\documentclass[showpacs]{revtex4-2}
\usepackage{graphicx}
\usepackage{dcolumn}
\usepackage{amsmath}
\usepackage[latin1]{inputenc}
\usepackage{graphicx, psfrag}
\usepackage{amssymb}
\usepackage[colorlinks=true, citecolor=blue, urlcolor = blue, linkcolor= red, bookmarks=true]{hyperref}
\usepackage{float}
\usepackage{amsmath}
\usepackage{amsfonts}
\usepackage{dcolumn}
\usepackage{hyperref}
\usepackage{subfigure}
\usepackage{pgfplots}
\usepackage{epstopdf}
\usepackage{booktabs}
\usepackage{siunitx}

\def \beq{\begin{equation}}
\def \eeq{\end{equation}}
\def \bse{\begin{subequations}}
\def \ese{\end{subequations}}
\def \bea{\begin{eqnarray}}
\def \eea{\end{eqnarray}}
\def \bem{\begin{displaymath}}
\def \eem{\end{displaymath}}
\def \bem{\begin{pmatrix}}
\def \eem{\end{pmatrix}}
\def \beb{\begin{bmatrix}}
\def \eeb{\end{bmatrix}}
\def \bc{\begin{center}}
\def \ec{\end{center}}

\def \nn{\nonumber}

\makeatletter
\def\btt#1{\texttt{\@backslashchar#1}}
\DeclareRobustCommand\bblash{\btt{\@backslashchar}} \makeatother

\makeatletter
\def\btt#1{\texttt{\@backslashchar#1}}
\DeclareRobustCommand\bblash{\btt{\@backslashchar}} \makeatother

\begin{document}
\title{Shadows of black hole surrounded by anisotropic fluid in Rastall theory}
\author{Rahul Kumar$^{a,b}$}\email{rahul.phy3@gmail.com}
\author{Balendra Pratap Singh$^{a}$}\email{balendra29@gmail.com}
\author{Md Sabir Ali$^{c}$}\email{alimd.sabir3@gmail.com}
\author{Sushant~G.~Ghosh$^{a,\;b}$} \email{sghosh2@jmi.ac.in, sgghosh@gmail.com}

\affiliation{$^{a}$ Centre for Theoretical Physics, Jamia Millia
Islamia, New Delhi 110025, India}
\affiliation{$^{b}$ Astrophysics Research Centre, School of Mathematics, Statistics and Computer Science, \\
University of KwaZulu-Natal, Private Bag 54001, Durban 4000, South Africa}
\affiliation{$^{c}$ Department of Physics, Indian Institute of
	Technology, Ropar, Rupnagar, Punjab 140001, India}

\begin{abstract}
Due to the gravitational lensing effect, a black hole casts a shadow larger than its horizon over a bright background, and the shape and size can be calculated.  The Event Horizon Telescope collaboration has produced the first direct image (shadow) of the black hole and it is in accordance with the shadow of a Kerr black hole of general relativity. However deviations from the Kerr black hole  arising from modified theories of gravity are not ruled out and they  are important as they offer an arena to test these theories through astrophysical observation.  This stimulates us to investigate rotating black holes surrounded by anisotropic fluid in Rastall theory namely a rotating Rastall black hole, which are characterized by mass $M$, spin $a$, field structure parameter $N_s$ and the Rastall parameter $\psi$. It  encompasses, as special cases, Kerr ($N_s \to 0$) and Kerr-Newman ($s=0$ and $N_s = -Q^2 $) black holes. The rotating Rastall black hole is characterized by an additional cosmological-like horizon apart from Cauchy and event horizons.   We derive an analytical formula for the shadow of a rotating Rastall black hole and go on to visualize the shadow of black holes for various values of the parameters for an observer at a given coordinates ($r_O, \theta_O$) in the domain $[r_+,r_q]$. 
\end{abstract}
\maketitle
\section{Introduction}
One of the possible extension of the general relativity (GR) is to relax the usual conservation law $\nabla_{\mu}T^{\mu\nu}=0$. Indeed it may not be always true in the curved spacetime \cite{Rastall:1973nw}. This motivated Rastall \cite{Rastall:1973nw} to proposed that the covariant divergence of energy$\textendash$momentum tensor may not be vanishing and should depends on the spacetime curvature through a coupling parameter, such that the GR (or divergenceless energy$\textendash$momentum tensor) is smoothly recover in the limit of zero coupling. He proposed that the divergence of the energy$\textendash$momentum tensor should be proportional to the gradient of the Ricci scalar. Thus the corresponding field equation of the Rastall theory reads
\begin{equation}
R_{\mu\nu}-\frac{1}{2}g_{\mu\nu}R=\kappa(T_{\mu\nu}-\lambda g_{\mu\nu}R).\label{eq001}
\end{equation}
Equation (\ref{eq001}) can be rewritten as
\begin{eqnarray}\label{Rastalleq}
R_{\mu\nu}+\left(\psi-\frac{1}{2}\right) g_{\mu\nu}R=\kappa T_{\mu\nu},
\end{eqnarray}
where $\kappa$ can be identified as gravitational coupling constant in the Rastall theory (hereafter $\psi=\kappa\lambda$, a Rastall coupling parameter).
It turns out that all vacuum solutions of GR are also solution of Rastall theory. However, the non-vacuum solutions depend upon the Rastall coupling parameter and are significantly different from corresponding solutions in GR, thereby making the Rastall gravity aesthetically rich \cite{Heydarzade:2017wxu}. 

Recently, Rastall theory of modified GR grabbed the great attention and a rich diversified research dedicated to it are available in literature including on standard cosmology \cite{Fabris:2011wz, Batista:2011nu,  Moradpour:2016rml}, loop quantum gravity \cite{Silva:2012gn},  Kaluza-Klein theory \cite{Wolf:1986wq}, and Brans-Dicke theory \cite{Smalley:1975ry}.
It is worthwhile mentioning here that the Rastall theory seems to be phenomenologically supported by particle creation process in the curved space time \cite{Bertlemann,Birrell:1982ix,Parker:1971pt}. The energy exchange between the spacetime geometry and the gravitating fluid accounts for this non-conservation of fluid's stress-energy tensor. However, little is known about Rastall gravity exact solutions, which deserve to be understood better to get insight of effect of non-minimal coupling between spacetime geometry and matter field. Nevertheless,
recently, interesting measures have been taken to get the
spherically symmetric black hole solutions of  Rastall gravity
\cite{Heydarzade:2016zof}. In particular, the static spherically symmetric black hole solutions with anisotropic fluid background has been found  by Heydarzade and Darabi in  Ref.~\cite{Heydarzade:2017wxu}. The line element has the following form 
\begin{equation}
ds^2=-f(r)dt^2+f(r)^{-1}dr^2+r^2(d\theta^2+\sin^2\theta d\phi^2), \label{eq0001}
\end{equation} 
where 
\begin{equation}
f(r)=1-\frac{2M}{r}-\frac{N_s}{r^{\frac{1+3\omega_s-6\psi(1+\omega_s)}{1-3\psi(1+\omega_s)}}},\label{fr}
\end{equation}
the surrounding fluid is characterize by the anisotropic energy-momentum tensor \cite{Visser:2019brz}. This study has also been extended to the noncommutative spacetime \cite{Ma:2017jko}, while the effect of Rastall parameter on black hole thermodynamics is shown in Refs.~\cite{Lobo:2017dib, Kumar:2017qws}. Also, some work compared Rastall theory with Einstein's GR in Refs.~\cite{Visser:2017gpz,Moradpour:2017tbp,Hansraj:2018zwl}. Even in the study of Neutron stars, the Rastall gravity is found to be consistent with the observations and could be used to placed constraints on the Rastall parameters \cite{Oliveira:2015lka}. More studies on Rastall gravity that demonstrate a significant deviation from the GR results can be found in the Refs. \cite{Li:2019jkv,Ziaie:2019jfl,Moradpour:2017ycq,Moradpour:2017shy,Yuan:2016pkz,Bronnikov:2016odv,Batista:2012hv,Batista:2011nu}. The no-hair theorem recommends the Kerr black hole as an astrophysical black hole candidate, but there still lacks direct conclusive evidence for this. Indeed, to the date, the so far observations could not provide a definite concrete proof of the Kerr nature of the astrophysical black hole, and it is still believe that astrophysical black holes may show deviation from the exact Kerr black holes \cite{Johannsen:2013rqa, Johannsen:2016uoh}. This opens an arena for investigating properties of black holes that differ from Kerr black holes. Johannsen and Psaltis \cite{Johannsen:2011dh} proposed a new model to test the no-hair theorem in the strong gravity regime using the rotating non-Kerr black hole metric characterized by number of deformation parameters. The generalization of stationary black holes in the
Rastall gravity to the axially symmetric case, Kerr-Newman (KN) like black hole, was
addressed recently \cite{Kumar:2017qws, Xu:2017bix}. Even though black holes have been immensely studied theoretically since their inception, the most conclusive evidence of supermassive black hole existence at the galactic center only came from the recent detection of M87* black hole shadow by the Event Horizon Telescope (EHT) collaboration \cite{1,Akiyama:2019cqa,Akiyama:2019brx,Akiyama:2019sww,Akiyama:2019bqs,Akiyama:2019fyp,Akiyama:2019eap}. The observed shadow of M87* black hole offers an unprecedented tool to test the strong gravity features of gravity, together it also test the assumption of the no-hair theorem and constrain the various black hole parameters (see e.g. \cite{Falcke:2013ola,Doeleman:2008xq}). Though the M87* black hole shadow is in accordance with the shadow of a Kerr black hole of GR, however, deviations from the Kerr black hole  arising from modified theories of gravity are not ruled out \cite{BambiM87}.\\
Rastall gravity is believed to have considerable deviation from the GR in high curvature regime, nevertheless, it would be interesting to examine whether it also has substantial effects on the horizon scale, where curvature is comparatively small. Interestingly, quantum effects originating in the very high curvature regime are shown to have significant impacts on the near horizon geometry \cite{Maldacena:2013xja, Giddings:2011ks}. Similarly, the noncommutativity \cite{Wei:2015dua}, asymptotic safety \cite{Held:2019xde,Kumar:2019ohr} and metric fluctuation \cite{Giddings:2016btb} are found to have observable effects on the black hole shadow. 
With these motivations, in this paper we investigate the shadow of a rotating black hole surrounded by an anisotropic fluid in the Rastall theory. In turn, we found that the surrounding field modified by the Rastall coupling indeed left a significant imprint on the black hole shadow. Since the shape and size of shadow unveil important properties of near-horizon spacetime, thus obtaining the shadow of rotating black hole in the presence of surrounding fluid and comparing it with observational results would provide us a best way to test the hypothesis of Rastall theory. The first ever motivation to study the black hole shadow was led by Synge \cite{Synge:1966} in 1966, who investigated the shadow of Schwarzschild black hole. Later, Bardeen \cite{Bardeen1} studied the Kerr black hole shadow. Since then a comprehensive study of shadow for a wide class of black holes including isolated rotating black holes as well as surrounded black hole have been made, i.e. Kerr-Newman black hole \cite{De}, black holes in extended Chern-Simons modified gravity \cite{Amarilla:2010zq}, Kaluza-Klein rotating dilaton black hole \cite{Amarilla:2013sj}, and regular black holes  \cite{Yumoto:2012kz,  Abdujabbarov:2016hnw, Amir:2016cen}, etc. This subject of interest has been extended to higher-dimensional spacetime by several researchers \cite{Papnoi:2014aaa,Abdujabbarov:2015rqa,Amir:2017slq,Singh:2017vfr}. In GR, the rotating black hole shadow is nearly circular, and due to the no-hair theorem its shape and size depend upon the mass and spin of black hole \cite{Bardeen1,Hioki:2009na}. However, for modified theories or black hole surrounded by some matter distributions, richer structures are possible \cite{Atamurotov:2015nra,Perlick:2015vta,Goddi:2016jrs,Takahashi:2005hy,Wei:2013kza,Abdujabbarov:2012bn,Amarilla:2011fx,Bambi:2008jg,Atamurotov:2013sca,Wang:2017hjl,Schee:2008kz,Grenzebach:2014fha,Cunha:2015yba,Cunha:2016bpi,Singh:2017xle, Abdujabbarov:2015pqp, Younsi:2016azx, Cunha:2018gql}. Over the past years, shadow has been investigated for some exotic alternatives to black holes \cite{Gyulchev:2018fmd,Shaikh:2018lcc,Shaikh:2018kfv,Gyulchev:2019tvk,Abdikamalov:2019ztb}. The study of black hole shadow is also a prominent way to estimate the characteristic parameters of black holes, i.e. the deviation of shadow from a circle is a measure for spin and other higher moments of the black hole \cite{Tsukamoto:2014tja, Hioki:2009na, Bambi:2010hf,Abdujabbarov:2015xqa,Kumar:2018ple}. Therefore, it is time to advance the theoretical investigations of the black hole shadow in view of the available shadow observational results.\\
The paper is organized as follow. In Sec. \ref{sect2}, we summarize the relevant properties of space-times of the rotating  black holes surrounded by an anisotropic fluid. Further in Sec.\ref{sect3}, we discuss the null geodesics and characterize the possible photon orbits around black hole. In Sec. \ref{sect4}, we determine the black hole shadow for an observer at finite distance from the black hole. Energy emission rate for black hole surrounded by photon sphere is discussed in Sec. \ref{sect5}. In Sec. \ref{sect6}, we summarize our main results.\\
The metric considered here, for definiteness, henceforth termed as a {\it rotating Rastall black hole}, a generalization of the Kerr metric as well as of the Kerr black hole surrounded by an anisotropic matter. We worked in four dimensional spacetime with metric signature (-, +, +, +) and use a system of units in which $c = G =\hbar= 1$. Greek indices are taken to run from 0 to 3.

\section{Rotating black hole in Rastall theory}\label{sect2} 
The vacuum rotating black holes in GR are described by the Kerr metric \cite{Kerr:1963ud}, which is completely defined by two parameters viz., mass and angular momentum. The static black hole in Rastall theory (\ref{eq0001}) \cite{Heydarzade:2017wxu} was extended to the Kerr-like black hole in Ref. \cite{Kumar:2017qws, Xu:2017bix}. Starting with the solution (\ref{eq0001}) and applying the Azreg-A\"{i}nou's \cite{Azreg-Ainou:2014pra,Azreg-Ainou:2014aqa} modified Newman-Janis algorithm, the rotating Rastall black hole metric is obtained, which is characterize by mass $M$, spin parameter $a$, surrounding fluid structure parameter $N_s$ and Rastall coupling parameter $\psi$ \cite{Xu:2017bix}. The metric for rotating Rastall black hole in the  Boyer$\textendash$Lindquist coordinates reads \cite{Xu:2017bix}
  
  \begin{equation}
  \begin{split}
  ds^2 =& - \left(1-\frac{2Mr+N_s r^{\zeta}}{\Sigma}\right)dt^2 - \frac{2a\sin^2\theta(2Mr + N_s r^{\zeta})}{\Sigma}d\phi dt+\Sigma d\theta^2+\frac{\Sigma}{\Delta}dr^2  \\
  &+ \sin^2\theta\left(r^2+a^2 +\frac{a^2\sin^2\theta(2Mr+N_s r^{\zeta})}{\Sigma} \right){d\phi}^2,
  \end{split} \label{RotMet}
  \end{equation}
with
\begin{eqnarray}
\Delta &=& r^{2}-2Mr+a^{2}-N_s r^{\zeta},\nonumber\\
{\Sigma}&=& r^2+a^2\cos^{2}{\theta},\nonumber\\
\zeta &=& \frac{1-3\omega_s}{1-3\psi(1+\omega_s)}.
\end{eqnarray}
In the limit $N_s\to 0$, the black hole metric (\ref{RotMet}) goes over to the usual Kerr black hole \cite{Kerr:1963ud}; $\omega_s$ is the state parameter of surrounding fluid. 
For the particular case $\psi=0$ and $-1< \omega_s<-1/3$, the metric represents the Kerr black hole surrounded by quintessence \cite{Ghosh:2015ovj}, while the Schwarzschild solution can be obtained in the limit when both $N_s, a\rightarrow0$.
The metric Eq.~(\ref{RotMet}) is clearly independent of $\phi$ and $t$ coordinates, and hence the time translation and rotational invariance of spacetime respectively infer the presence of two Killing vectors $\eta_t^{\mu}=\delta^{\mu}_t$ and $\eta_{\phi}^{\mu}=\delta^{\mu}_{\phi}$. 

The  horizons structure  of the rotating Rastall black hole is quite different from the GR counterpart. Depending upon the black hole mass $M$, spin parameter $a$, structure parameter $N_s$ and the suitable choice of Rastall coupling constant $\psi$ the spacetime may admits upto three horizons, viz., Cauchy horizon, event horizon and quintessential (or cosmological) horizon (depending upon surrounding field) located at radial coordinates $r_-, r_+, r_q$, respectively.  A numerical analysis of horizon radii are shown in Table \ref{table1} and \ref{table2}. 
It turns out that, for a given $a$, there exists a critical value of $\psi$ (=$\psi^E$) such that $\Delta=0$ has a double root which corresponds
to an extremal black hole with degenerate horizons ($r_+=r_q\equiv r^E$) as shown in Table \ref{table1} and \ref{table2}. When $\psi<\psi^E$, $\Delta=0$ has three simple zeros, while it has only one zero for $\psi>\psi^E$. These two cases correspond, respectively, to a non-extremal black hole with a Cauchy horizon, event horizon and quintessential horizon, and a spacetime  with only one horizon.  Similarly, we can also find extremal black hole for $N_s^E$ (cf. Table \ref{table2}).
This can be easily noticed that event horizon radius increases while quintessential horizon radius decreases for the increasing values of Rastall coupling parameter ($\psi$) or field structure parameter ($N_s$). The higher the value of $\psi$ or $ N_s$, the more close the two outer horizons. Consequently, with the increasing Rastall coupling, the size of black hole event horizon grows. Apart from the horizons, rotating black hole also exhibit another surface of physical significance; the static limit surface whose radial coordinates are the solution of $g_{tt}=0$ \cite{Kumar:2017qws, Xu:2017bix}. The peculiarity of Rastall theory is that fluid's stress tensor $T_{\mu\nu}$ (not conserved) get modifies owing to a non-zero Rastall coupling ($\psi$) and behaves as an effective field of conserved stress tensor $T'_{\mu\nu}$, thus the solutions of modified Einstein's field equation in this theory are effectively different from those of GR \cite{Moradpour:2017tbp}, i.e. a dust field with $\omega_s=0$ may get modified to a quintessence or to an even stronger repulsive phantom field, subject to the suitable choice of $\psi$. The non-zero value of $\psi$ introduces a new character to the black hole and thus the spacetime properties near a black hole would be drastically changed. 
\begin{table}
 \centering
\begin{tabular}{ c c  c  c  c c}
 \hline
  $\psi$   & $r_-$ &  $r_+$ &  $r_{q}$\\
 \hline
    $-0.07$   & 0.133872  & 2.06354 &38.9333\\
    \hline
    $0.0$ & 0.133905  &2.10198 &17.7641\\
  \hline 
 $ 0.05$ & 0.133924 & 2.14202 &11.2483  \\
  \hline
  $ 0.08$& 0.133934 & 2.1746 &8.79995\\
  \hline
   $0.10$ &0.133939   & 2.2018 &7.53879\\
  \hline
   $(\psi)^{E} $ &0.133961 & 2.93053 & 2.93053\\
  \hline
   $0.25 $ & - & - & 0.133965\\
  \hline
  
  \end{tabular}
    \caption{The horizon radii (Cauchy $r_-$, event $r_+$, quintessential horizon $r_q$) of rotating Rastall black holes for various coupling parameter $\psi$ and black hole parameters $M=1$, $a=0.5$, and field structure parameter $N_s=0.05$. Degenerate horizon $(r_+=r_q= 2.93053)$ occur for extremal value of $\psi^E=0.21070669$.}
  \label{table1}
\end{table}
\begin{table}
 \centering
\begin{tabular}{ c c  c  c  c c}
 \hline
  $N_s$   & $r_-$ &  $r_+$ &  $r_{q}$\\
 \hline
    $0.01$   & 0.133964  & 1.90945 &51.5794\\
    \hline
    $0.03$ & 0.133944  &2.0112 &18.7598\\
  \hline 
 $ 0.05$ & 0.133924 & 2.14202 &11.2483  \\
  \hline
  $ 0.07$& 0.133904 &2.32462 &7.70627\\
  \hline
   $0.08$ &0.133894   & 2.45243 &6.49261\\
   \hline
   $N_s^{E}$ &0.133868&3.53436&3.53436\\
  \hline
  $0.12$ &-&-&0.133854\\
  \hline
  \end{tabular}
    \caption{The horizon radii of rotating Rastall black holes  ($\omega=-2/3$)  for varying parameter $N_s$ with $\psi=0.05$. Degenerate horizon $(r_+=r_q=3.53436)$ occur for extremal value of $N_s^{E}=0.1052697428252$. }
  \label{table2}
\end{table}
\section{Photons geodesics around rotating black hole \label{sect3}}
Let us consider that the black hole is in the front of an extended source of light. Depending upon the photon energy and angular momentum these source photons may get scatter, capture or move in the unstable orbits \cite{Chandrasekhar:1992}.  The marginally trapped photons which revolve around the black hole many times before reaching to the distant observer form black hole shadow boundary. Tracing back these photons trajectory will give us the apparent shape and size of black hole shadow, which, for the Kerr black hole, depends upon the black hole parameters ($M, a$) and interestingly on the inclination angle ($\theta_O$) between the direction of a distant observer and axis of rotation. Therefore, in order to discuss the black hole shadow, we study the null geodesics for metric (\ref{RotMet}). We adopt the Hamilton-Jacobi equation and Carter constant separable method \cite{Carter:1968rr} to study the complete geodesic equation of motion.
The most general form of Hamilton-Jacobi equation reads:
\begin{eqnarray}
\label{HmaJam}
\frac{\partial S}{\partial \tau} = -\frac{1}{2}g^{\alpha\beta}\frac{\partial S}{\partial x^\alpha}\frac{\partial S}{\partial x^\beta} ,
\end{eqnarray}
where $S=S(\tau,x^\mu)$ is the Jacobean action and $\tau$ is the affine parameter. Now, for stationary and axially symmetric spacetime Eq.~(\ref{RotMet}) we are taking a separable solution for the Jacobi action $S$ as \cite{Chandrasekhar:1992}
\begin{eqnarray}
S=\frac12 {m_0}^2 \tau -{\cal E} t +{\cal L} \phi +S_r(r)+S_\theta(\theta) \label{action},
\end{eqnarray}
The energy $\mathcal{E}$ and angular momentum $\mathcal {L}$ are two conserved quantities corresponding to the time translational and rotational invariance, respectively; apart from the conserved norm of four-velocity. In addition to these obvious constants of motion, Carter \cite{Carter:1968rr} used the separability of the Hamilton-Jacobi equation and explicitly showed the existence of new conserved quantity in terms of Carter constant. This ensures the complete integrability of geodesic equations. We follow the standard variable separable method and obtain the complete equations of motion around the rotating black hole Eq.~(\ref{RotMet}) in the first-order form, which takes the following form: 
\begin{eqnarray}
\Sigma \frac{dt}{d\tau}&=&\frac{r^2+a^2}{\Delta}\left({\cal E}(r^2+a^2)-a{\cal L}\right)  -a(a{\cal E}\sin^2\theta-{\mathcal {L}})\ ,\label{tuch}\\
\Sigma \frac{dr}{d\tau}&=&\pm\sqrt{\mathcal {\mathcal{R}}}\ ,\label{r}\\
\Sigma \frac{d\theta}{d\tau}&=&\pm\sqrt{\Theta}\ ,\label{th}\\
\Sigma \frac{d\phi}{d\tau}&=&\frac{a}{\Delta}\left({\cal E}(r^2+a^2)-a{\cal L}\right)-\left(a{\cal E}-\frac{{\cal L}}{\sin^2\theta}\right)\ ,\label{phiuch}
\end{eqnarray}
where the expressions for 
$\mathcal{R} (r)$ and ${\Theta}(\theta)$ in Eq.~(\ref{r}) and (\ref{th}) takes the form  
\begin{eqnarray}\label{06}
\mathcal{R}(r)&=&\left[(r^2+a^2){\cal E}-a{\cal L}\right]^2-(r^2-2 M r+a^2-N_s r^{\zeta})[{m_0}^2r^2+(a{\cal E}-{\cal L})^2+{\cal K}],\quad \\ 
\Theta(\theta)&=&{\cal K}-\left[\frac{{\cal L}^2}{\sin^2\theta}-a^2 {\cal E}^2\right]\cos^2\theta,
\end{eqnarray}
with $\mathcal{K}$ as the Carter constant. The geodesic of a test particle around rotating Rastall black hole are completely determined by the equations (\ref{tuch})-(\ref{phiuch}) and  for $N_s=0$, these geodesic equations reduce to the corresponding equations around Kerr black hole. To analyze the photons orbit we introduce two dimensionless impact parameters $\eta$ and $\xi$, which have a functional form in terms of energy $\mathcal{E}$, angular momentum $\mathcal{L}$ and Carter constant $\mathcal{K}$ as
\begin{equation}
\xi=\mathcal{L}/\mathcal{E}, \quad\quad \eta=\mathcal{K}/\mathcal{E}^2.
\end{equation}
Every orbit is characterize by the constants of motion  $\eta$ and $\xi$. Indeed,  $\eta$ and $\xi$ are further related to the celestial coordinates. 
Photons have zero rest mass ($m_0=0$) and follow null geodesics, for them the Eq.~(\ref{r}) in terms of dimensionless quantities $\eta$ and $\xi$ takes the form
\begin{equation}
\mathcal{R}(r)=\frac{1}{{\cal{E}}^2}\left[\left((r^2+a^2) -a{\xi}\right)^2-(r^{2}-2Mr+a^{2}-N_s r^{\zeta})\left((a -{\xi})^2+{\eta}\right)\right].
\end{equation}
The apparent shape of the black hole shadow is dictated by the null geodesics. It must be stressed out that geodesics are possible only when $(\mathcal{R}\geq0)$ (cf. Eq. (\ref{r})), therefore the nature of roots of $\mathcal{R}$ will distinguish the geodesics around the black hole. Out of all possible motion, e.g., capturing (no real root of $\mathcal{R}$), scattering ($\mathcal{R}$ has real roots for $r\geq r_+$), and unstable (circular and spherical) orbits; only the unstable orbits play a role in casting silhouette of black hole whose radii can be found by $(\mathcal{R}=d\mathcal{R}/dr=0)$. 
It must be noticed that shadow boundary is not the unstable photon orbits rather it is the \textit{apparent} shape of the unstable photons orbits, owing to the strong gravitational field in the vicinity of black hole causing the gravitational lensing.
In order to obtain the circular orbits of the photons, it is important to study the radial motion. One can rewrite the radial equation of motion (\ref{r}) as \cite{Wei:2013kza}
\begin{equation}
\left(\frac{d{r}}{d\tau}\right)^2+V_{eff}(r)=0,
\end{equation}
where $V_{eff}$ is the effective potential experienced by the test particle, which reads in terms of impact parameters
\begin{equation}
V_{eff}=-\frac{1}{{{\Sigma}^2\mathcal{E}^2}}\left[\left((r^2+a^2) -a{\xi}\right)^2-(r^{2}-2Mr+a^{2}-N_s r^{\zeta})\left((a -{\xi})^2+{\eta}\right)\right].\label{vef}
\end{equation}
The unstable orbits witness a radial turning point ($\dot{r}= \ddot{r}=0$) and occur for the maximum value of the effective potential
\begin{equation}
V_{eff}=\frac{\partial V_{eff}}{\partial r}=0 \;\; \;\; \mbox{or}\;\;  \; \mathcal{R}=\frac{\partial \mathcal{R}}{\partial r}=0.\label{vr} 
\end{equation}
These orbits are planer only at the equatorial plane, on the other hand, depending upon the value of Carter constant, a generic unstable orbit lies in a three dimensional plane and usually known as spherical photon orbits. Solving Eq.~(\ref{vr}) for impact parameters, yield
\begin{eqnarray}
\xi &=& \frac{1}{{r_p}^{\frac{3}{\sigma}} \left(a^2 \left(N_s (3 \omega_s -1){r_p}^{\frac{3 \omega_s +1}{\sigma}}+2\sigma (M-{r_p}) {r_p}^{\frac{2}{\sigma}+1}\right)\right)}\Big[ a^3 N_s (3 \omega_s -1) {r_p}^{\frac{3 \omega_s +4}{\sigma}} \nonumber\\
&-& a N_s(2\sigma-3\omega_s+1) {r_p}^{\frac{2\sigma+3\omega_s+4}{\sigma}}- 2 a \sigma {r_p}^{\frac{4}{\sigma}+2} \left(N_s{r_p}^{\frac{3 \omega_s }{\sigma}}-\left(a^2+{r_p} ({r_p}-2 M)\right) {r_p}^{\frac{1}{\sigma}}\right)\nonumber \\
&+& 2 a M (a^2-{r_p}^2)\sigma {r_p}^{\frac{5}{\sigma}+1}\Big], \label{xiexp}\\
\eta &=&\frac{-{r_p}^{2-\frac{2}{\sigma }}}{a^3 \left(N_s (3 \omega_s -1) {r_p}^{\frac{3 \omega_s +1}{\sigma }}+2 \sigma  (M-r) {r_p}^{\frac{2}{\sigma }+1}\right)^2}\Big[-4N_s a^3 {\sigma}(2 \sigma- 3 \omega_s + 1) {r_p}^{(\frac{2\sigma + 3\omega_s +5}{\sigma})} \nonumber \\
&+&  4 a \sigma {r_p}^{\frac{6}{\sigma }+3} \left(N_s {r_p}^{\frac{3 \omega_s-1 }{\sigma }}- \left(a^2+{r_p} ({r_p}-2 M)\right)\right) \left(N_s (2 \sigma - 3 \omega_s +1) {r_p}^{\frac{3 \omega_s-1 }{\sigma }-1}+2 M \sigma \right) \nonumber \\
&+&  a N_s {r_p}^{(\frac{3 \sigma + 3 \omega_s+ 5}{\sigma})}\left(4 M \sigma (6 \sigma - 3 \omega_s + 1) +N_s \left( (2\sigma-3\omega_s+1)^2 +4\sigma^2 \right) {r_p}^{\frac{3 \omega_s-1 }{\sigma }-1}\right)
   \nonumber \\
&+& 4 a \sigma ^2 \left(5 M^2-4 M {r_p}+{r_p}^2\right) {r_p}^{\frac{6}{\sigma }+4}-8 Ma^3{\sigma}^2 {r_p}^{(3 +\frac{6}{\sigma})}  -8 a N_s \sigma ^2 {r_p}^{(\frac{4 \sigma + 3 \omega_s+ 5)}{\sigma})}\Big], \label{etaexp}
\end{eqnarray}
where $\sigma=3\psi(1+\omega_s)-1$ and $r_p$ is the unstable orbit radius. For the non-rotating black hole, these unstable orbits make a photon sphere of constant radius around the black hole, whereas, for the rotating black hole, photons will either have motion along black hole rotation or against its rotation. The radii of these prograde ($r_p^-$) and retrograde ($r_p^+$) orbits are different which lead to the formation of photon region around rotating black hole i.e., a photon region is filled with unstable photon orbits. For instance, unstable circular photon orbit radius around Kerr black hole has range $M\leq r_p\leq 4M$, whereas for rotating Rastall black hole it depends upon various parameters $N_s, \omega_s, \psi$. The expressions of $\xi$ and $\eta$ in  Eqs.~(\ref{xiexp}) and (\ref{etaexp}) are sufficient to determine the contour of the shadow in the ($\xi, \eta$) plane, which revert back for the standard Kerr black hole in the absence of the surrounding field ($N_s=0$) \cite{Hioki:2009na}.
\begin{eqnarray}
\xi&=&\frac{a^2 (M+{r_p})+{r_p}^2 \left({r_p}-3 M\right)}{a (M-{r_p})},\ \\
\eta &=&\frac{{r_p}^3 \left(4 a^2 M -{r_p}\left({r_p}-3 M\right)^2\right)}{a^2 (M-{r_p})^2}.
\end{eqnarray} 
These equations further determines the shadow of Schwarzschild black hole satisfying the relation $\eta+\xi^2=27M^2$, for $a=\psi=N_s=0$ and $r_p=3M$.
\section{Shadow of rotating Black hole in Rastall gravity \label{sect4}}
In this section, we will discuss the black hole shadow more elaborately and see whether the Rastall gravity marks some significant imprint on a shadow. A black hole shadow helps to understand the properties of black hole spacetime and provide a strong tentative way to find the parameters associated with a black hole. The photons moving in unstable orbits around a black hole form shadow silhouette which appears as a sharp boundary between bright and dark region. In ideal settings the shape and size of shadow completely depend upon the black hole parameters and spacetime geometry \cite{Falcke:2000, Chandrasekhar:1992}. However, it is expected that its shape and size will get influenced by the gravity governing a black hole and the matter around it. 

The left portion of shadow is constructed by prograde photons, whereas the right portion by retrograde photons (assuming the counter-clockwise rotation of black hole) \cite{Hioki:2009na}. For a rotating black hole, photons moving on prograde and retrograde orbits experience different effective potential such that the prograde orbits form closer to the horizon than the retrograde ones, and hence the prograde photons that have reached the observer, are visible from a lesser angular distance from the black hole \cite{Bambi:2008jg}. Furthermore, the radius of prograde orbit decreases whereas it increases for a retrograde orbit with increasing spin parameter of the black hole, which results in the noticeable off-center shifting of shadow. For large values of the spin parameter ($a\gtrsim 0.5$), the change in radius is higher for prograde orbits than the retrograde one, which appears as a squeezing of shadow in the direction perpendicular to black hole angular momentum. As a result, the Kerr-like shadow appears as a distorted disk when compared with a shadow of a non-rotating black hole which is a perfect circular disk \cite{Bardeen1, Chandrasekhar:1992}.
Here, we are considering the surrounding field as an anisotropic fluid, namely, quintessence, and hence, the rotating Rastall black hole is no longer asymptotically flat and a cosmological horizon encompassing the event horizon is present at a finite distance due to an anisotropic fluid (cf. Table \ref{table1}, \ref{table2}). The region between event horizon ($r_+$) and cosmological horizon ($r_q$) is of particular interest because in this region $g_{rr}$ is positive and vector field $\partial_r$ is spacelike. Overstepping this region $r<r_+$ or $r>r_q$, change the causal character of vector field $\partial_r$ from spacelike to timelike, and no static observer can be considered in the said region. The observer outside the $r_q$ will be driven away from the black hole due to cosmic expansion, whereas inside $r_+$ observer will be inevitably falling toward singularity due to immense gravitational pull \cite{Perlick:2018iye}. This prevents to consider a static shadow observer at spatial infinity or at an arbitrarily large distance. Furthermore, we shall restrict our discussion on the equatorial plane ($\theta_O=\pi/2$)  as the observer on this plane witnesses the immense variation of the black hole shadow with varying rotation parameter. Henceforth, we will be considering a static observer placed at a radial coordinate $r_O$ such that $r_+<r_O<r_q$, i.e., we consider an observer placed at Boyer-Lindquist coordinates ($t_O,r_O,\theta_O=\pi/2,\phi_O=0$). For an observer at $r_O$ ($r_+<r_O<r_q$), the event horizon at $r_+$ is future inner horizon while cosmological horizon at $r_q$ is future outer horizon, and crossing any of these two horizons will break the causal connection with region $r_+<r<r_q$. For a study of the light motion around the black hole, we are considering a spreaded light source present behind the black hole in the region $r_p<r<r_q$ \cite{Grenzebach:2014fha}, and we shall focus only on the unstable photon orbits. These orbits are unstable in the sense that any radial perturbation to them forces photons to leave either of the two future horizons. Rotating black hole exhibit two distinct photon regions filled with unstable orbits at $r>r_+$ and $r<r_-$, however, for shadow construction only the exterior photon region ($r>r_+$) is essential \cite{Grenzebach:2014fha}. \\
For shadow study, we consider a light ray originated from the observer position and back trace it into the past, then either it reaches to a point of the background source or crosses the horizon at $r_+$ and seemingly come from the
white hole. We see the brightness in the direction of the first type of light rays and darkness in the direction of the second type. A black hole shadow is constructed by the borderline case of these two types i.e., light rays which undergo unstable motion around black holes and form orbits. In order to define the celestial coordinates along the shadow boundary, and to study the light sent back into the past, let construct the orthonormal basis at observer position. Following \cite{Grenzebach:2014fha}, we consider orthonormal tetrad
\begin{eqnarray}
e_0=\frac{(r^2+a^2)\partial_t+a\partial_{\phi}}{\sqrt{\Sigma\Delta}},\;\;\;\;e_1=\frac{1}{\sqrt{\Sigma}}\partial_{\theta},\nn\\
e_2=-\frac{\partial_{\phi}+a\sin^2{\theta}\partial_t}{\sqrt{\Sigma}\sin\theta},\;\;\;\; e_3=-\sqrt{\frac{\Delta}{\Sigma}}\partial_r.
\end{eqnarray} 

\begin{figure*}[!ht]
    \begin{tabular}{c c c c}
	\includegraphics[scale=0.67]{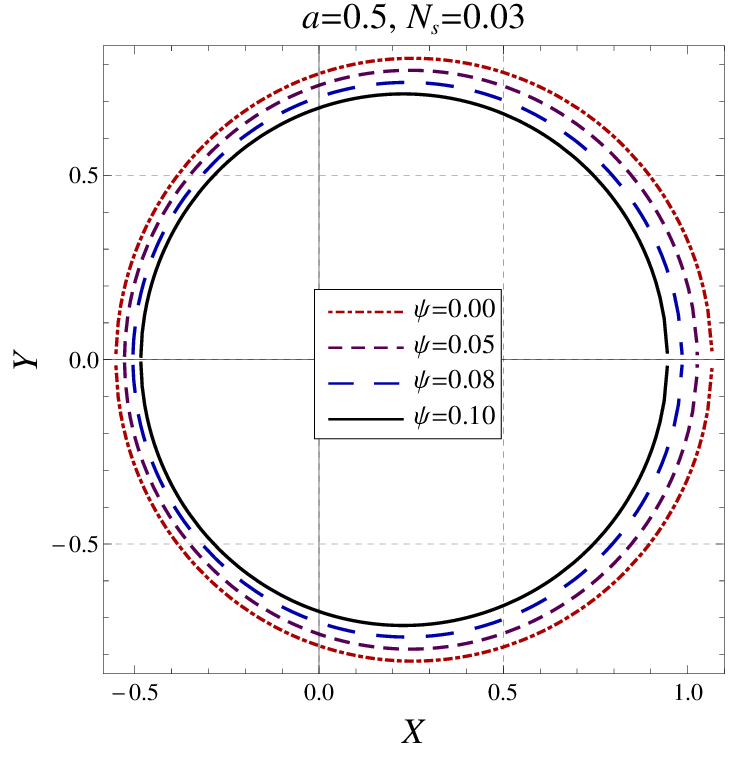} 
	\includegraphics[scale=0.63]{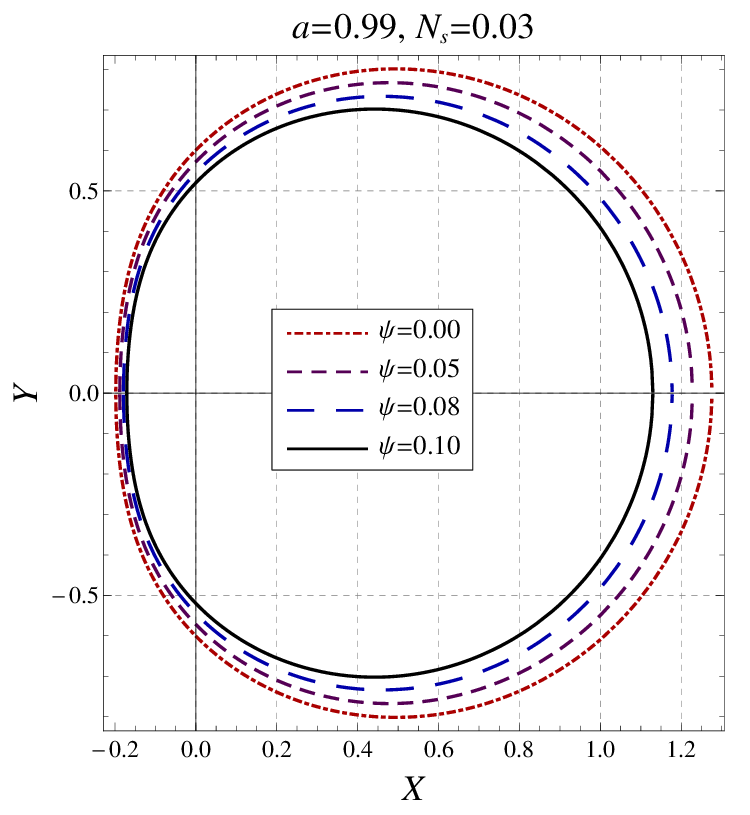} \\
    \includegraphics[scale=0.67]{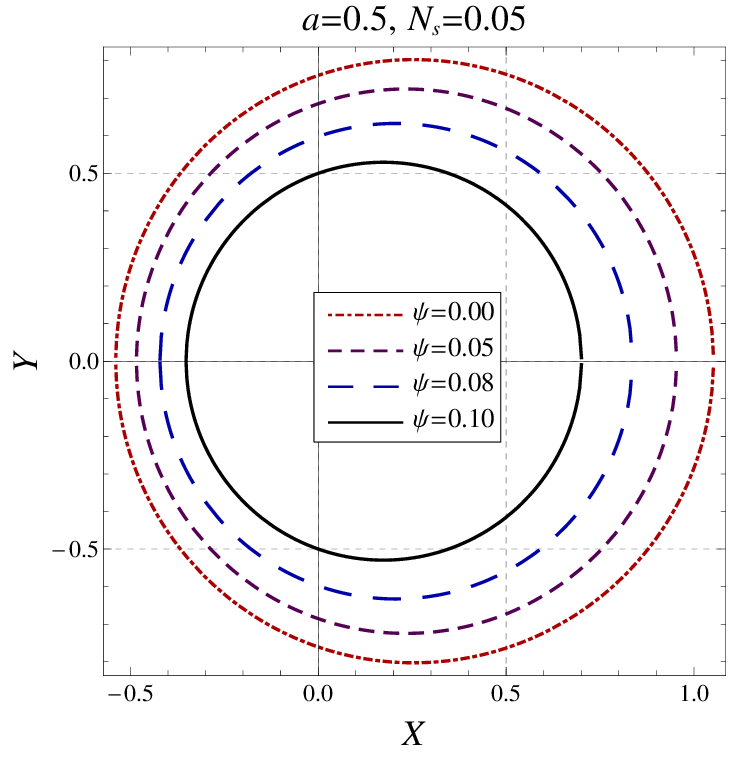}
	\includegraphics[scale=0.63]{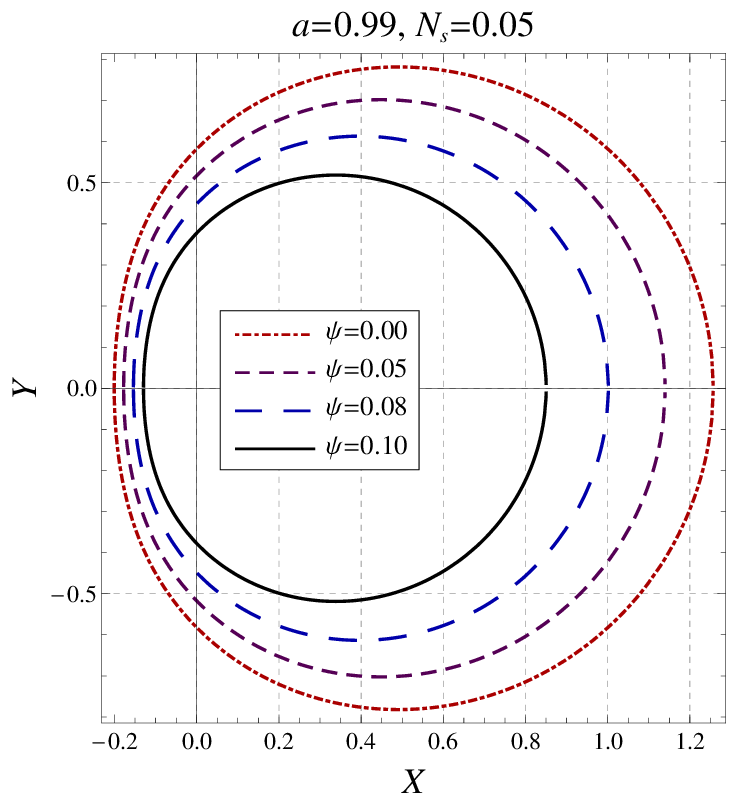} \\
	 \end{tabular}
   \caption{\label{fig2} Rotating Rastall black hole shadows for different values of $a$, $N_s$ and varying $\psi$. In all plots outer curve (\textit{dashed dotted red}) corresponds to the shadow of Kerr black hole surrounded by an anisotropic fluid ($\psi=0, \omega_s=-2/3$). The shadow corresponds to each curve and the region inside it.}
\end{figure*}

\begin{figure*}[!ht]
    \begin{tabular}{c c c c}
	\includegraphics[scale=0.68]{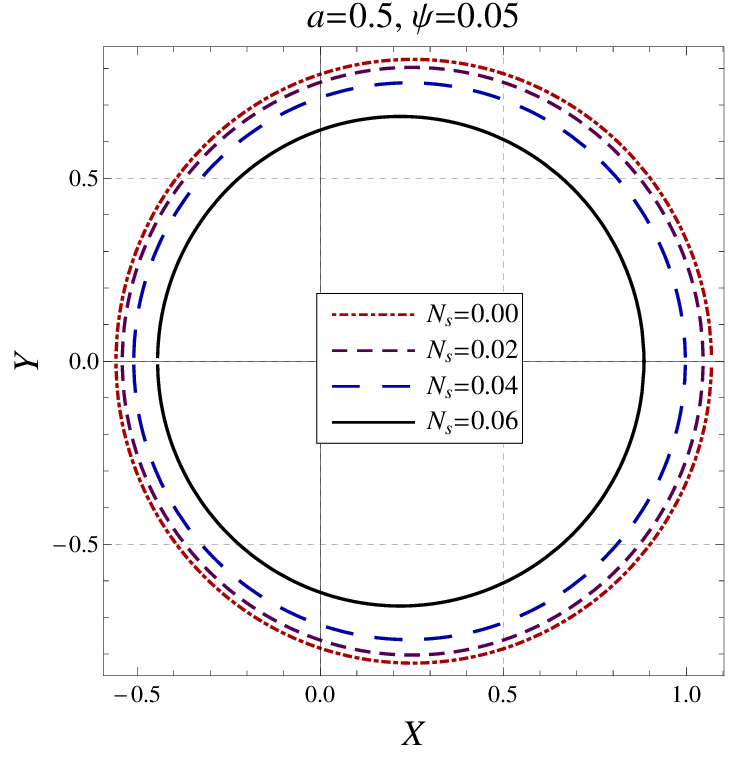} 
	\includegraphics[scale=0.68]{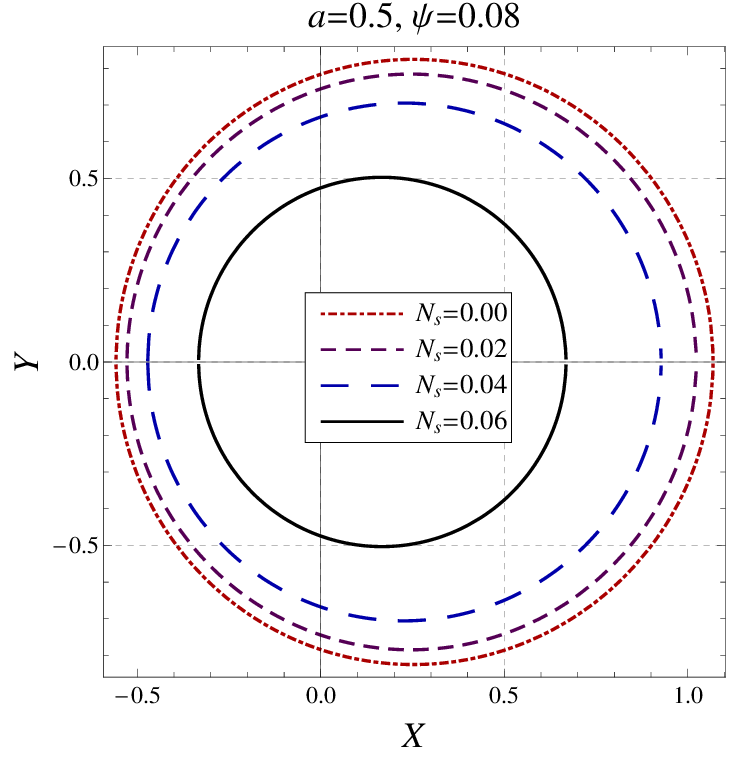} \\
    \includegraphics[scale=0.68]{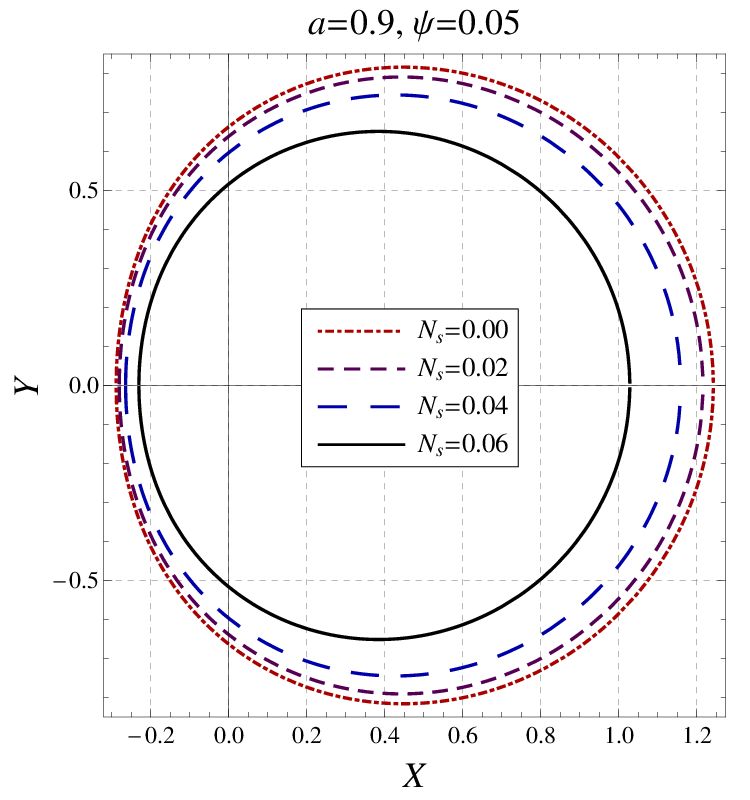}
	\includegraphics[scale=0.68]{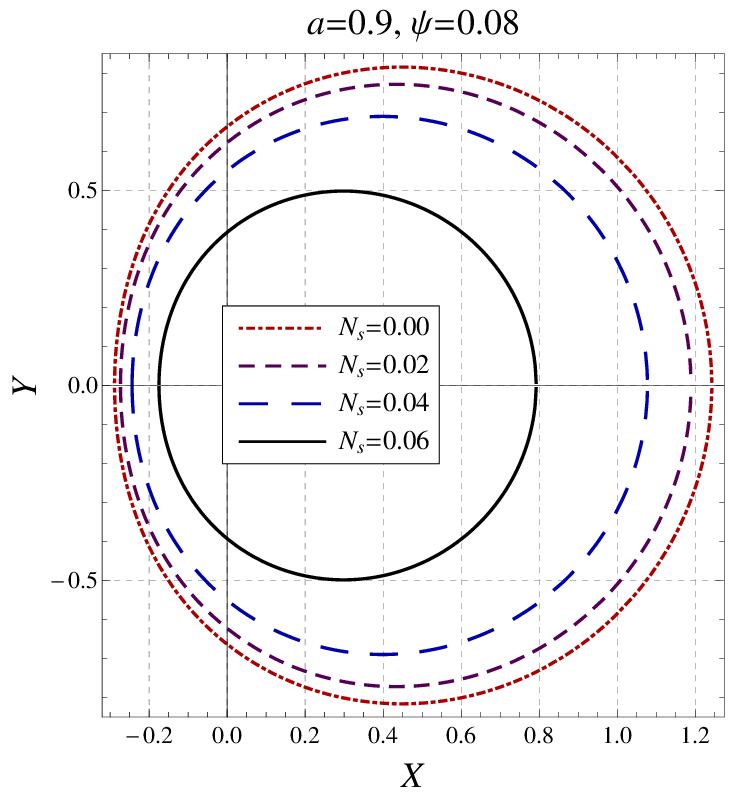}\\
	 \end{tabular}
        \caption{\label{fig3}  Rotating Rastall black hole shadows for different values of $a$, $\psi$ and with varying $N_s$. The outer curve (\textit{dashed dotted red}) corresponds to the Kerr black hole shadow ($N_s=0$). The shadow corresponds to each curve and the region inside it.}
\end{figure*}

\begin{figure*}[!ht]
	\begin{tabular}{c c c c}
		\includegraphics[scale=0.62]{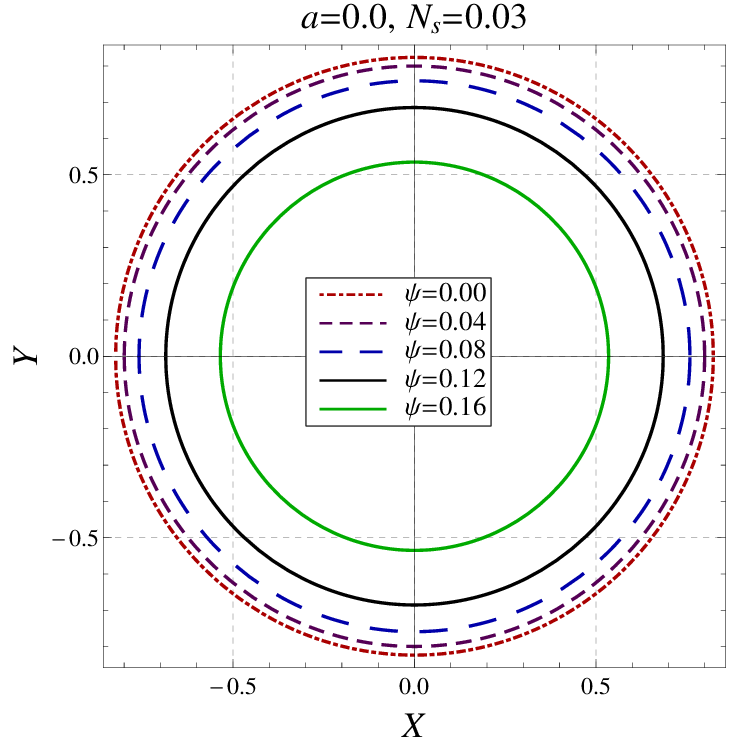} 
		\includegraphics[scale=0.62]{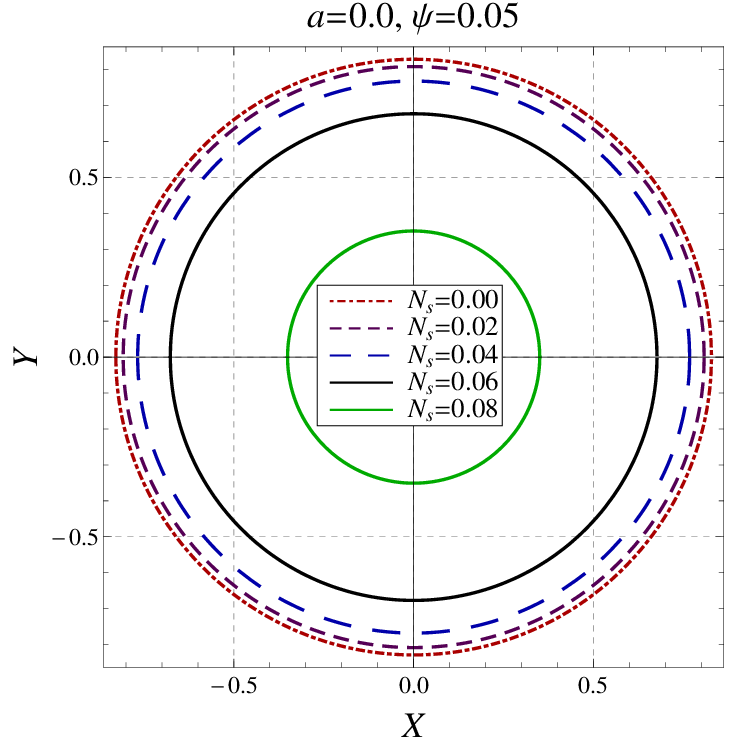} 
	\end{tabular}
	\caption{\label{SchShadow} The non-rotating Rastall black hole shadow with varying $\psi$ and $N_s$. In the left panel, outer (\textit{dashed dotted red}) curve corresponds to the Schwarzschild black hole shadow surrounded by quintessence ($\psi=0, \omega_s=-2/3$.) In the right panel, outer (\textit{dashed dotted red}) curve corresponds to the Schwarzschild black hole shadow.}
\end{figure*}

The coefficients of tetrad are real only if the observer is in the region $r_+<r_O<r_q$, as $\Delta$ is positive only in this region. A tangent vector to a light ray $\lambda(\tau)$ with affine parameter $\tau$ is
\begin{equation}
\dot{\lambda}(\tau)=\dot{t}\partial_t+\dot{r}\partial_r+\dot{\theta}\partial_{\theta}+\dot{\phi}\partial_{\phi},\label{lambda1}
\end{equation}
which at an observer position is \cite{Grenzebach:2014fha}
\begin{equation}
\dot{\lambda}(\tau)=\alpha(-e_0+\sin\Phi\cos\Psi e_1 +\sin\Phi\sin\Psi e_2 +\cos\Phi e_3 ).\label{lambda2}
\end{equation} 
The scalar factor $\alpha$ is determined by comparing Eq. ($\ref{lambda1}$) and (\ref{lambda2}), and using Eqs. (\ref{tuch}) and (\ref{phiuch}), which reads
\begin{equation}
\alpha=\frac{a\mathcal{L}-(r^2+a^2)\mathcal{E}}{\sqrt{\Sigma\Delta}}.\label{alpha1}
\end{equation}
Now, the celestial coordinates $\Phi$ and $\Psi$ can also be calculated as \cite{Grenzebach:2014fha}
\begin{eqnarray}
\sin\Psi(r_O,r_p) &=&\left.\left( \frac{\xi-a}{\sqrt{(a-\xi)^2+\eta}}\right)\right|_{r=r_O},\\ \label{Psieq}
\sin\Phi(r_O,r_p) &=&\left.\left(\frac{\sqrt{\Delta[(a-\xi)^2+\eta]}}{(r^2+a^2-a\xi)}\right)\right|_{r=r_O},\label{Phieq}
\end{eqnarray}
with $\xi$ and $\eta$ are function of photon unstable radius. To visualize a shadow we introduce the Cartesian coordinates $X$ and $Y$ \cite{Grenzebach:2014fha}
\begin{eqnarray}
X(r_O,r_p)=-2\tan\left(\frac{\Phi(r_O,r_p)}{2}\right)\sin(\Psi(r_O,r_p)), \label{X11}\nn\\
Y(r_O,r_p)=-2\tan\left(\frac{\Phi(r_O,r_p)}{2}\right)\cos(\Psi(r_O,r_p)).\label{Y1}
\end{eqnarray}  
A plot of $X$ vs $Y$ shall gives the boundary of a shadow for the rotating Rastall black hole as seen by an static observer at $r_O$. We assume that observer is in the region $r_+< r_O< r_q$ and choose the spacetime parameter appropriately. The shadow of rotating Rastall black holes for different values of the spin parameter $a$ and field parameter $N_s$ with varying Rastall coupling constant $\psi$ is depicted for $\omega_s=-2/3$ in Fig.~\ref{fig2}. The observer is placed at $r_O=6M$ in the region $r_+<r_O<r_q$. For a comparative study, the shadow of a Kerr black hole surrounded by quintessence field ($\psi=0$) is also shown in all plots. We find that the rotating Rastall black hole shadow size decrease with increasing value of parameter $\psi$. In Fig.~\ref{fig3}, the rotating Rastall black hole shadows with varying field structure parameter $N_s$ and different values of parameters $a$ and $\psi$ are shown. In all plots, the shadow of Kerr black hole is also shown. The size of black hole shadow monotonically decreases with increasing $N_s$. However, for a large value of $\psi$, the effect of $N_s$ is more prominent and shadow size decreases more rapidly with increasing $N_s$. The sufficiently large values of $N_s$ make the shadow apparently more symmetric along the $Y$-axis and subsided the generic asymmetry caused by the spin of black hole. Most importantly, increasing values of both parameters $\psi$ and $N_s$ individually decreases the size of the shadow. The shadow for non-rotating Rastall black hole is shown in Fig.~\ref{SchShadow}, which is as expected a perfect circle. The shadow size decrease with increasing $\psi$ or $N_s$. 

\begin{figure}[!ht]
\begin{tabular}{ c c}
\includegraphics[scale=0.65]{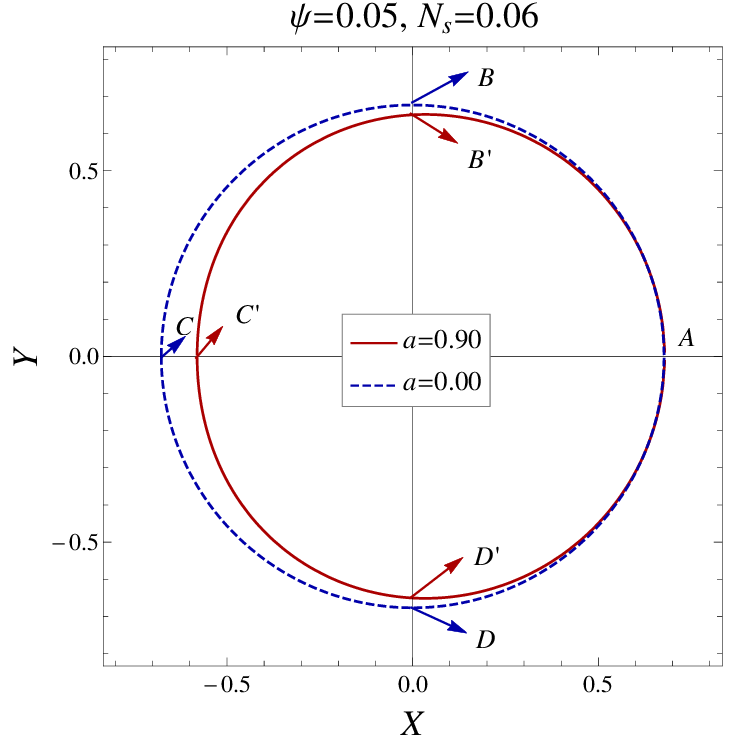}
\includegraphics[scale=0.65]{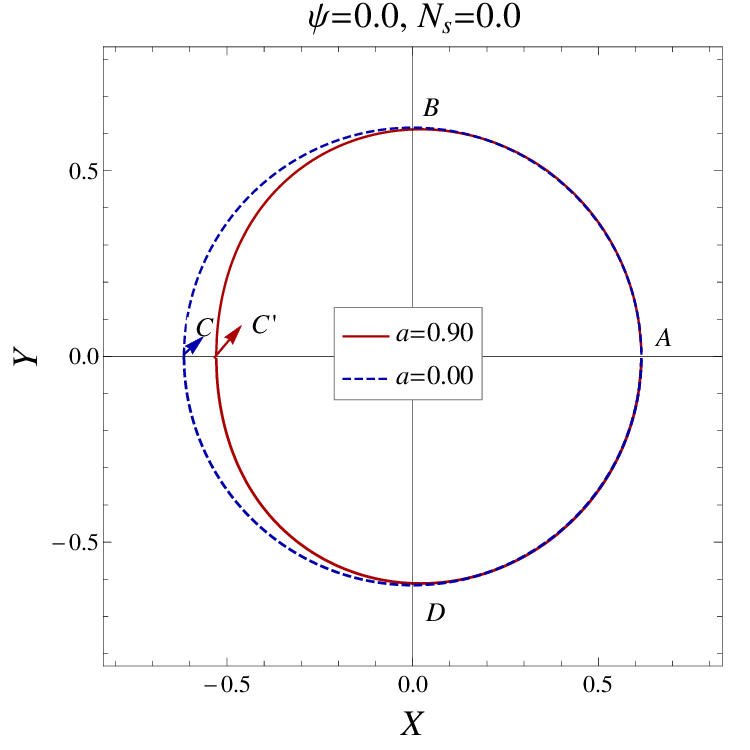}
\end{tabular}

\caption{The comparison of rotating and non-rotating black hole shadow (\textit{Left panel}) in Rastall gravity $\psi=0.05$, $N_s=0.06$, (\textit{Right panel}) Schwarzschild and Kerr black hole shadow  $\psi=N_s=0.0$. The shadow silhouette of rotating black hole is represent by (\textit{red solid curve}) and for non-rotating black hole by (\textit{blue dashed curve}). Shadow of rotating black hole is shifted along $X-$axis.   }\label{obs2}
\end{figure}
\begin{figure}[h!]
    \begin{tabular}{c c c c}
	\includegraphics[scale=0.66]{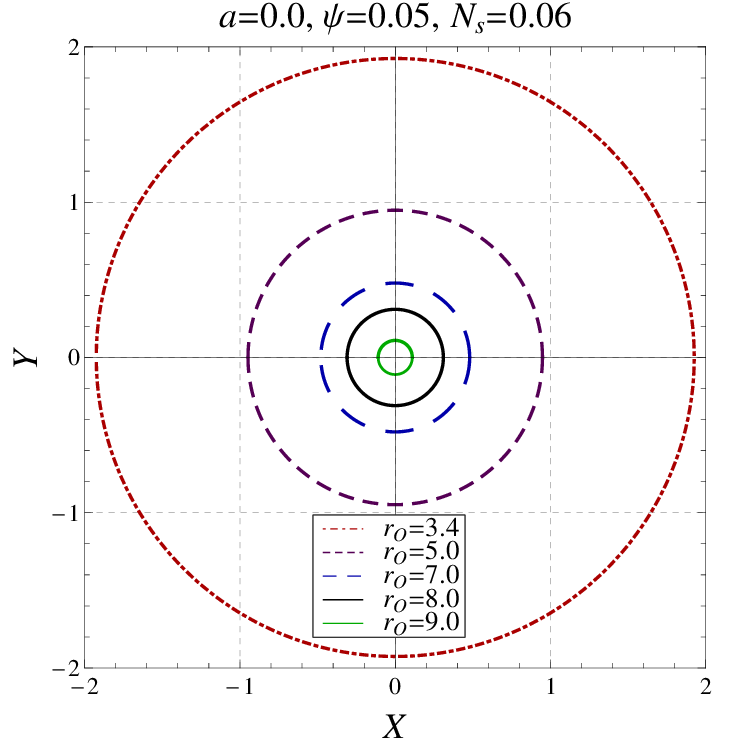} 
	\includegraphics[scale=0.66]{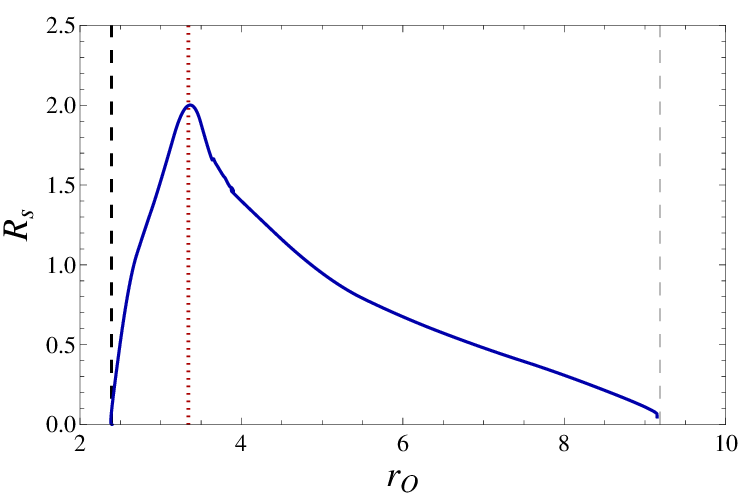} 
   	\end{tabular}
    \caption{\label{fig4} (\textit{Left panel}) The non-rotating Rastall black hole shadows for varying observer position $r_O$, where $r_+=2.395$, $r_q=9.186$. (\textit{Right panel}) Variation of black hole shadow radius with observer position; black dashed lines represent the horizon positions and red dotted line represents the radial coordinate of photon sphere $r_p=3.34121$. Shadow radius increases as observer move from $r_q$ to $r_p$, whereas the radius of bright apparent Universe decrease as observer move further from $r_p$ to $r_+$, see text for details. }
\end{figure}

Next, we compare the non-rotating and rotating black hole shadows in Rastall theory with those for Schwarzschild and Kerr black hole in Fig.~\ref{obs2}, and found that the shadows of non-rotating and rotating Rastall black holes cut the coordinate axis $Y$ at different points ($B, D$) and ($B', D'$), respectively, whereas Schwarzschild and Kerr black hole shadows cut it at same point (cf. Fig. \ref{obs2}). Interestingly, for particular values of $\psi$ and $N_s$ the shadow diameter of rotating Rastall black hole along the $Y$-axis as perceive by an observer at $r_O$, is smaller than the corresponding non-rotating black hole shadow (cf.~Fig~\ref{obs2}), unlike in GR where the shadow diameter remains insensitive to black hole spin. Conclusively, the presence of non-zero Rastall coupling makes a significant impact on the black hole shadow and Fig.~\ref{fig2} and \ref{obs2} are compellingly inferring that the black hole shadow in the Rastall theory is noticeably different from the corresponding shadow in GR.\\
For the non-rotating black hole, the variation of shadow radius with observer position in the region $r_+\leq r_O\leq r_q$ is shown in Fig.~\ref{fig4}. Shadow radius increases monotonically as the observer move from the cosmological horizon ($r_q$) to the photon sphere ($r_p$). Due to the presence of ergoregion and frame dragging around rotating black hole, we can not consider a static observer in the immediate vicinity of event horizon. However, for the non-rotating black hole it is easy to descend the observer upto the event horizon. Although, it may seem physically irrelevant for astrophysical purposes to study the black hole shadow for an observer inside the photon region or very close to the event horizon. Nevertheless, it would always be informative and interesting to study such cases.
Consider an observer placed at ($r_O,\theta_O=\pi/2$) emits a flash of light rays pointing toward black hole, then those rays which makes an angle in the range $[-\vartheta,\vartheta]$ will fall into the black hole and account for the dark region of shadow, where angle is measured with respect to radial line joining observer and black hole. This angle reads as  \cite{Bozza:2009yw}

\begin{equation}
\vartheta=\arcsin\left(\frac{\xi}{r_O}\sqrt{f(r_O)}\right)=\arcsin\left(\frac{\xi}{r_O}\sqrt{1-\frac{2M}{r_O}-\frac{N_s}{{r_O}^{\frac{1+3\omega_s-6\psi(1+\omega_s)}{1-3\psi(1+\omega_s)}}}}\right).
\end{equation}
Accordingly  $\vartheta$ gives the angular radius of shadow and $\pi-\vartheta$ is the escape cone angle, which clearly depends on the critical impact parameter for photons, and observer position.
\begin{figure}[b!]
\includegraphics[scale=0.35]{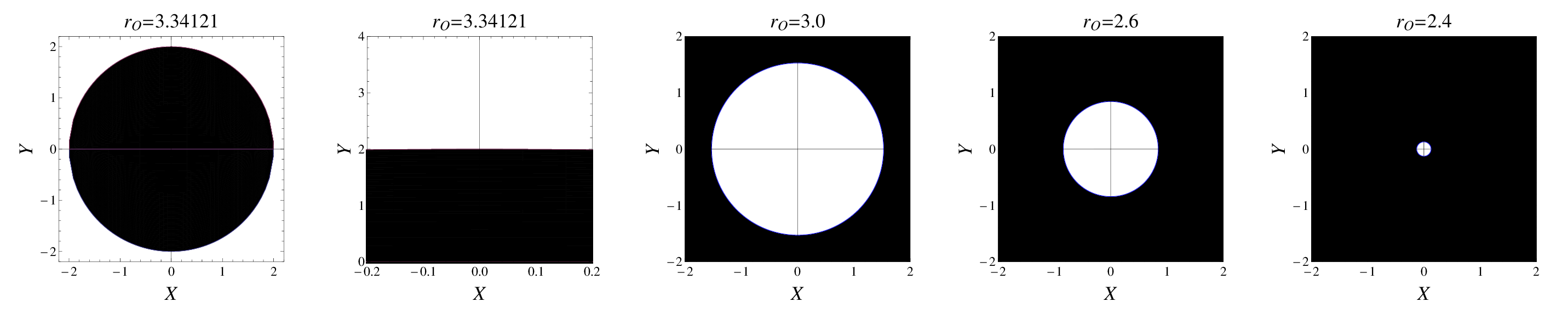}
\caption{The non-rotating black hole shadow in Rastall gravity with varying $r_O$ for $\psi=0.05$ and $N_s=0.06$. The $r_+=2.3950,\;  r_p=3.34121, \; r_q=9.18625$. The viewing angle of observer from looking directly at the black hole are, respectively, $\ang{0}$, $\ang{90}$, $\ang{180}$, $\ang{180}$, $\ang{180}$. The white region is the bright visibal Universe. } \label{new15}
\end{figure}
For instance, if we take the particular values $N_s=0.06, \psi=0.05, \omega_s=-2/3$.
Then for an observer at cosmological horizon ($r_O=r_q=9.18625$), angle is $\vartheta=0$, i.e., the angular radius of black hole shadow is zero, and hence the observer sky will be completely bright. On the other hand, for an observer at the boundary of photon sphere ($r_O=r_p=3.34121$), $\vartheta=\pi/2$, and hence the observer's half sky $[-\pi/2,\pi/2]$ is dark while other half is bright. Whereas for observer at event horizon $r_O=r_+=2.3950$, the angle is $\vartheta=\pi$ which means that event horizon take over the entire frontal field of view and makes the observer's entire sky completely dark $[-\pi,\pi]$ \cite{Perlick:2018iye, Nemiroff:1993he}. In essence, the angular size of black hole shadow increases monotonically and observer's sky turns from completely bright to completely dark as observer moves from cosmological horizon to the event horizon. An observer, located exactly at the photon sphere, see complete darkness while looking toward black hole, half bright and half dark sky while looking toward the tangential plane of photon region, and a bright sky as looking opposite to the black hole.  When an observer descend to the photon sphere, the escape cone for light rays start getting shrinking and ultimately disappear at $r_+$. Therefore, for an observer inside the photon sphere entire bright sky will be limited in the shape of a circular disk, whose angular radius ($\pi-\vartheta$) decrease monotonically on approaching the event horizon of black hole, and finally completely vanishes at the horizon (cf. Fig. \ref{new15}). Indeed, for an observer at infinitesimally close to event horizon of black hole, the only light rays emanating at perfectly radially outward direction will be able to escape the black hole, whereas those rays fired at any other angle will spiral around black hole and will eventually fall into the singularity. In Fig.~\ref{new15}, we have shown the shadow for an observer outside and inside the photon sphere.
\begin{figure}[t!]
	\begin{tabular}{c c c }
		\includegraphics[scale=0.67]{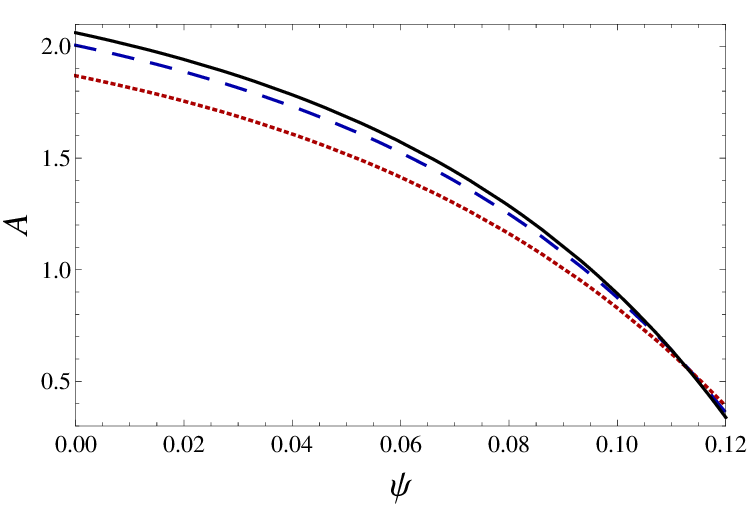} \\
		\includegraphics[scale=0.67]{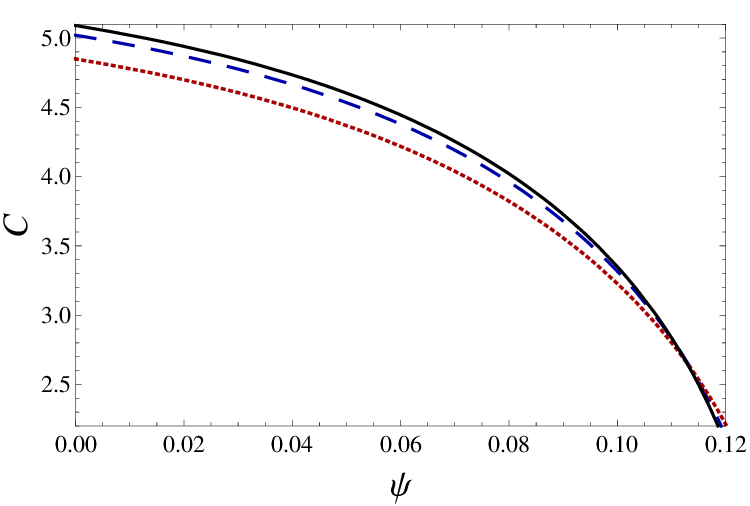} \\
		\includegraphics[scale=0.67]{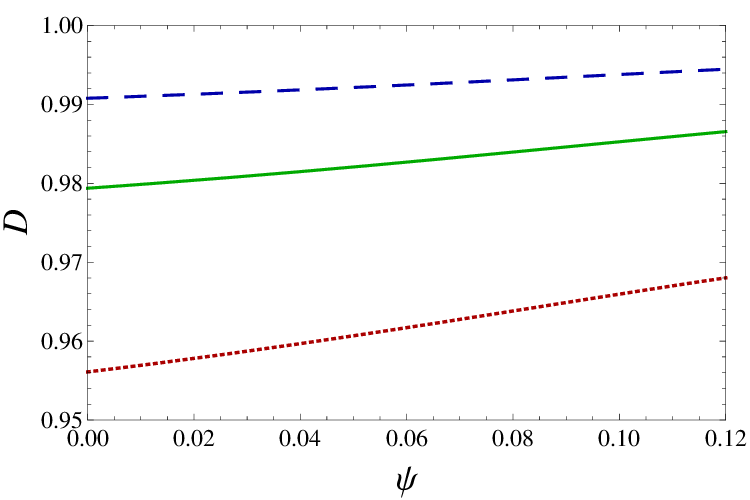} 
	\end{tabular}
	\caption{\label{Rastall} The plot of area $A$, circumference $C$ and oblateness $D$ for rotating Rastall black hole  ($N_s=0.05, \omega_s=-2/3$) with varying coupling parameter is shown. (\textit{Solid black curve}) for $a=0.0$, (\textit{Dashed blue curve}) for $a=0.5$, (\textit{Solid green curve}) for $a=0.7$ and (\textit{Dotted red curve}) for $a=0.9$.}
\end{figure}
\begin{figure*}
	\begin{tabular}{c c c }
		\includegraphics[scale=0.7]{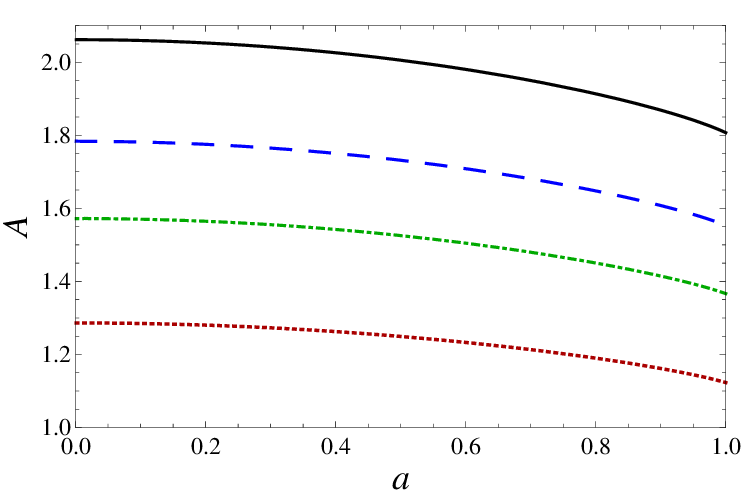} \\
		\includegraphics[scale=0.7]{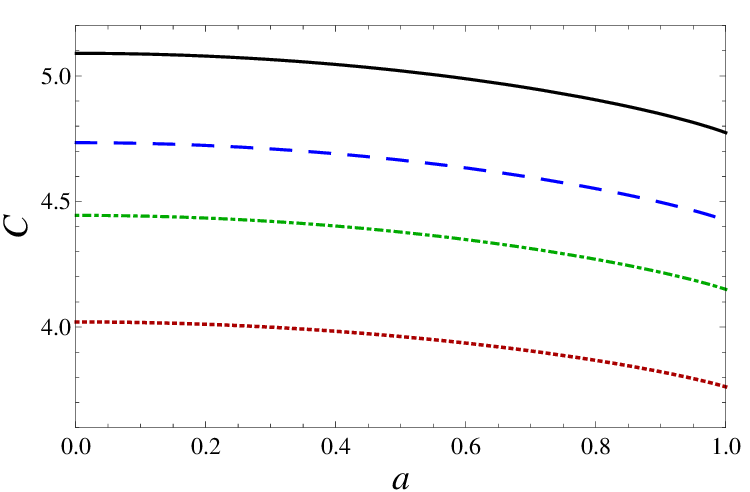} \\
		\includegraphics[scale=0.7]{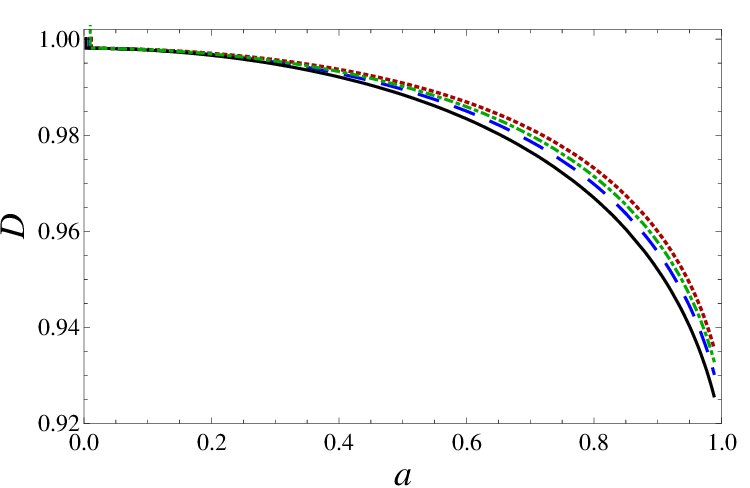} 
	\end{tabular}
	\caption{ The plot of area $A$, circumference $C$ and oblateness $D$ for rotating Rastall black hole surrounded ($N_s=0.05, \omega_s=-2/3$) with varying spin parameter is shown. (\textit{Solid black curve}) for $\psi=0.0$, (\textit{dashed blue curve}) for $\psi=0.04$, (\textit{dotted dashed green curve}) for $\psi=0.06$ and (\textit{dotted red curve}) for $\psi=0.08$.}\label{Rastall001}
\end{figure*}
We introduce astronomical observables namely, shadow area $A$, its circumference $C$ and oblateness $D$ to characterize the shadow size and shape \cite{Kumar:2018ple}
\begin{eqnarray}
A &=&2\int{Y(r_p) dX(r_p)}=2\int_{r_p^{-}}^{r_p^+}\left( Y(r_p) \frac{dX(r_p)}{dr_p}\right)dr_p,\nonumber\\
C &=& 2\int\sqrt{({dY(r_p)}^2+{dX(r_p)}^2)}=2\int_{r_p^{-}}^{r_p^+}\sqrt{\left(\left(\frac{dY(r_p)}{dr_p}\right)^2+\left(\frac{dX(r_p)}{dr_p}\right)^2\right)}dr_p,\nonumber\\
D&=&\frac{X_r-X_l}{Y_t-Y_b},\label{Observables}
\end{eqnarray}
where the subscript $r, l, t$, and $b$, respectively, stand for right, left, top, and bottom of shadow boundary. $A$ and $C$ characterize the size, whereas $D$, the ratio of horizontal and vertical diameters, measure the distortion in shadow and characterize its shape.
In Fig.~\ref{Rastall} and \ref{Rastall001}, we have shown these shadow observables for $r_O=6M$ and varying $\psi$ and $a$. $A$ and $C$ both decrease with increasing Rastall coupling parameter $\psi$ or spin parameter $a$. However, $\psi$ and $a$ play relatively opposite role in the distortion of shadow as $\psi$ make the shadow more symmetric along $Y-$axis viz. the value of oblateness parameter increases with $\psi$ but decrease with $a$ (cf. Fig.~\ref{Rastall} and \ref{Rastall001}). In Fig. \ref{Rastall01}, we plotted $A$ and $D$ in the ($a$, $\psi$) plane, the point where the contours of $A$ and $D$ intersects, give the value of black hole spin parameter and Rastall coupling parameter. Hence, if one measure the area and oblateness of shadow through observation, the spin and Rastall coupling parameter could be easily determined, e.g., if $A=1.4$ and $D=0.98$ then $a=0.7426$ and $\psi=0.06566$.

\begin{figure*}[b!]
	\begin{tabular}{c c}
	\includegraphics[scale=0.9]{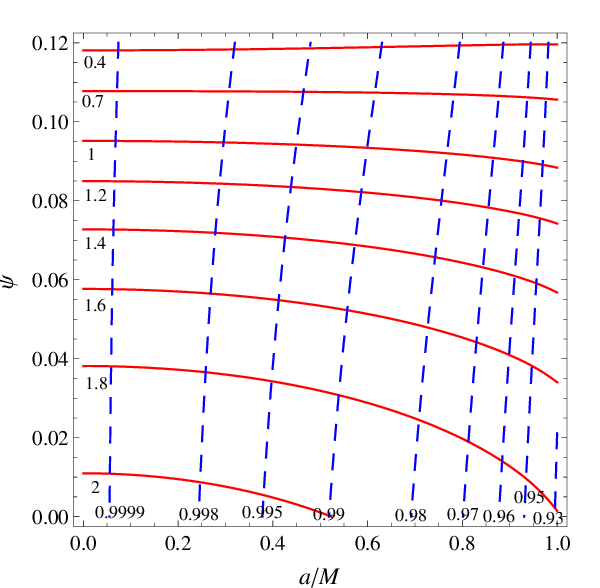} &
	\includegraphics[scale=0.9]{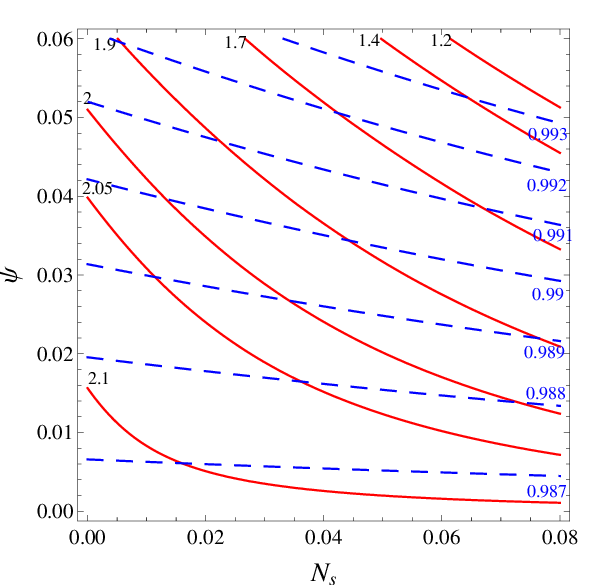} 
	\end{tabular}
	\caption{Contour plots of the observables $A$ and $D$ in the plane $(a, \psi)$ for Rastall black hole with $N_s=0.05$ (\textit{Left Panel}) and in the plane $(\psi, N_s)$ with $a=0.5$ (\textit{Right Panel}). Each curve is labeled with the corresponding value of $A$ and $D$. (\textit{Solid red curve}) corresponds to the area curve, and (\textit{Dashed blue curve}) for oblateness.}\label{Rastall01}
\end{figure*}

\section{Energy Emission Rate}\label{sect5}
Here, we will discuss the rate of energy emission for a black hole (\ref{RotMet}) surrounded by anisotropic fluid in Rastall gravity. The low energy absorption cross-section, for a spherically symmetric black hole, has a universal feature that it always reduces to the black hole horizon area \cite{Das:1996we}. However, at high energy scale the absorption cross section oscillate around a limiting constant value, which took the value of geometrical cross section ($\sigma_{lim}$) of the photon sphere in which the black hole is endowed  \cite{Decanini:2011xi}, the value is
\begin{equation}
 \sigma_{lim}\approx  \pi R_s^2, 
  \end{equation} 
where $R_s$ is black hole shadow radius.
  \begin{figure*}[!ht]
     \begin{tabular}{c c c c}
 \includegraphics[scale=0.7]{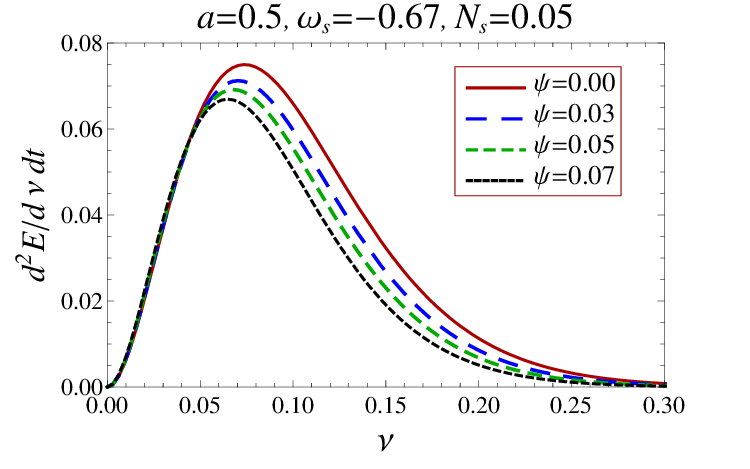}
 \includegraphics[scale=0.7]{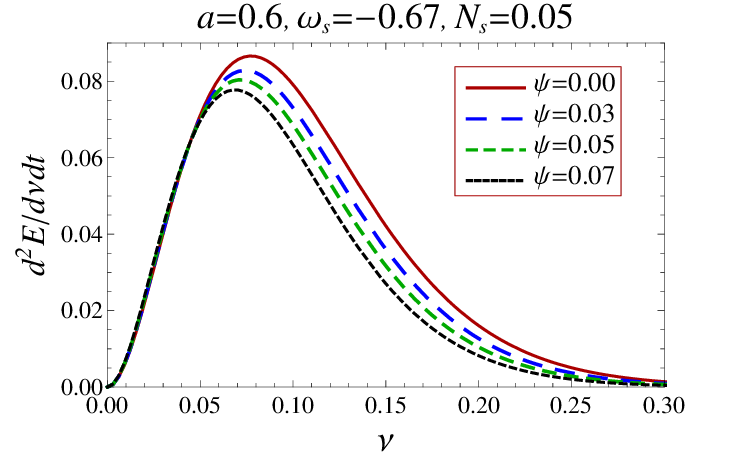}
 \end{tabular}
 \caption{\label{fig5}  The variation of energy emission rate  in the Rastall gravity with frequency $\nu$ for different values of $\psi$. }
 \end{figure*}
It is believed that it is a generic feature of black hole at high energy surrounded by photon sphere. 
The expression of the energy emission rate of black hole reads \cite{Mashhoon:1973zz,Misner:1973,Wei:2013kza}
  \begin{equation}
 \frac{d^2E(\nu)}{d\nu dt}= \frac{2 \pi^2 \sigma_{lim}}{\exp{(\nu/T_+)}-1}\nu^3,
 \end{equation}
 where $T_+$ is the Hawking temperature for the outer-most horizon ($r_+$) reads as \cite{Kumar:2017qws}
 \beq
 T_+=\frac{r_+^2-a^2+N_s(1-\zeta)r_+^\zeta}{4\pi r_+(r_+^2+a^2)}.
 \eeq
 In the absence of the surrounding field $(N_s=0)$ the expression of the black hole temperature reduces to
 \beq
 T_+=\frac{r_+^2-a^2}{4\pi r_+(r_+^2+a^2)}.
 \eeq
which is the temperature for the rotating Kerr black hole. In Fig.~\ref{fig5}, we have plotted energy emission rate for different values of $\psi$, $a$, $\omega_s$ and $N_s$ and it is clearly shown in figure (cf. Fig.{~\ref{fig5}}) that the Gaussian peak gets down with the increasing value of  $\psi$.
\section{Conclusion}\label{sect6}
The apparent boundary which separates photons geodesics that are trapped from those that can escape from the incredible gravitational pull of black hole is called the black hole shadow, and it appear as a dark region deprived of photons against a bright background. The ground-breaking observations by EHT collaboration provided the first direct evidence of black hole existence and its shadow. This offer a unique opportunity to probe gravity in its strongest regime and determining the exact nature of astrophysical black hole. \\
The rotating Rastall black holes, inheriting the non-minimal coupling between gravity and matter fields, belong to the family of non-Kerr black holes where $\Delta$ can be dependent on $\theta$ and $r$ both as shown in Ref.~\cite{Kumar:2017qws}. These black holes are important as it is believed that an astrophysical black hole may deviate from Kerr solution \cite{Johannsen:2013rqa, Johannsen:2016uoh}. It is found in general studies that non-rotating black hole shadow is symmetric along celestial coordinate axis, however the symmetry along the black hole rotational axis vanishes for rapidly rotating black hole, owing to the spin and angular momentum dependent effective potential for photon which eventually causes effectively different capture radius for prograde and retrograde photons. Hence, we made a qualitative analysis of shadow of black hole  surrounded by an anisotropic fluid in the Rastall theory.
 
The rotating Rastall black hole (\ref{RotMet}) is not asymptotically flat for suitable choices of parameters and a cosmological horizon naturally arises due to surrounding matter. This prevent to consider a static observer at arbitrary large distance, and compel to restrict it within the region enclosed by event and cosmological horizons. With this setup, we study the black hole shadow observed by an observer at radial coordinate $r_O$, such that $r_+< r_O<r_q$. We have investigated how the size and apparent shape of the black hole shadow is changed due to the Rastall parameter by analyzing unstable orbits. Although, the black hole solution is mathematically complicated, we could derive analytical formulas for the shadows of such black holes. We consider the particular case of surrounding anisotropic fluid  of state parameter $(-1\leq \omega_s\leq -1/3)$. It is found that the spin parameter and Rastall coupling parameter play significantly different roles in casting the shadow of rotating black hole. The spin parameter is effectively shifting the shadow to the right side in the celestial coordinate frame and sufficiently high values cause a dent on the left side of shadow. Whereas, for suitably higher values of $\psi$ and increasing $N_s$ the rotating black hole shadow become apparently more symmetric along the rotational axes.    
Furthermore, the size of shadow decreases with increasing Rastall parameter $\psi$  as well as with $N_s$, i.e., for a given value of rotational parameter $a$, the rotating Rastall black hole leads to a smaller shadow than the Kerr black hole.

The non-rotating black hole shadow is studied for an observer with varying position ranging from $r_+$ to $r_q$. It is found that, the shadow completely vanishes leaving a completely bright sky for an observer at the cosmological horizon, whereas, at the boundary of photon sphere $r_p$ the observer sky is half bright and half dark.

\section{Acknowledgments}
S.G.G. would like to thanks Department of Science and Technology for the INDO-SA bilateral project DST/INT/South Africa/P-06/2016, SERB-DST for the ASEAN project IMRC/AISTDF/CRD/2018/000042 and also IUCAA, Pune for the hospitality while this work was being done. R.K. thanks the NRF and the University of KwaZulu-Natal for continued support.

\end{document}